\documentclass{webofc}
\usepackage[varg]{txfonts}   
\usepackage{bm}
\begin{document}
\title{Gluon decay into heavy quark pair under a strong magnetic field}
%
%

\author{\firstname{Shile} \lastname{Chen}\inst{1}\fnsep\thanks{\email{csl18@tsinghua.org.cn}} \and
 \firstname{Jiaxing} \lastname{Zhao}\inst{2,3} \and
        \firstname{Pengfei} \lastname{Zhuang}\inst{1}
}
\institute{Physics Department, Tsinghua University, Beijing 100084, China
\and
Helmholtz Research Academy Hesse for FAIR (HFHF), GSI Helmholtz Center for Heavy Ion Physics, Campus Frankfurt, 60438 Frankfurt, Germany
\and
Institut f\"ur Theoretische Physik, Johann Wolfgang Goethe-Universit\"at,Max-von-Laue-Straße 1, D-60438 Frankfurt am Main, Germany}

\abstract{%
Due to the extreme large magnetic field produced in the initial stage of non-central heavy-ion collision, the dynamical process of gluon decay into heavy quark pair will take place under an external field rather than in vacuum.  Unlike in the vacuum case, where the process is forbidden by energy momentum conservation, under the external field, a process emerges considering the background energy which recovers the conservation. We calculate the gluon decay rate at leading order under a uniform magnetic field.}
\maketitle
%
It is widely accepted that the strongest electromagnetic field in nature is generated in high energy nuclear collisions~\cite{Kharzeev:2007jp,Voronyuk:2011jd,Deng:2012pc,Tuchin:2013ie}. The maximum of the magnetic field can reach $5m_{\pi}^2\sim 0.1\ \rm GeV^2$ in Au+Au collisions at top RHIC energy and almost $70m_\pi^2\sim 1\ \rm GeV^2$ in Pb+Pb collisions at LHC energy~\cite{Deng:2012pc,Tuchin:2013ie}, where $m_\pi$ is the pion mass. The strong magnetic field will cause many novel phenomena like chiral magnetic effect(CME)~\cite{Kharzeev:2007jp}, photoproduction of dileptons and quarkonia in peripheral and ultra-peripheral collisions~\cite{ALICE:2015mzu}, splitting of $D^0$ and $\bar D^0$ directed flows~\cite{Das:2016cwd}, and spin-polarized difference between $\Lambda$ and $\bar \Lambda$~\cite{Guo:2019joy}. 

In our previous study~\cite{Chen:2024lmp}, we calculate the heavy quark production with the elementary process $gg\to Q\bar Q$ under the extreme strong magnetic field and show the analytical results with the Lowest Landau Level approximation (LLL). Because of the dimension reduction in phase space, the cross section of this process has a divergence around the incoming energy threshold. Then the heavy quark pair production in nucleus-nucleus collisions will be extremely enhanced at low transverse momentum and suppressed at high transverse momentum. The existence of external magnetic field which breaks the rotational invariance and only the momentum along the magnetic field is conserved. The production process becomes anisotropic and strongly dependent on the direction of motion.

\begin{figure}[h]
\centering
\includegraphics[width=4.5cm]{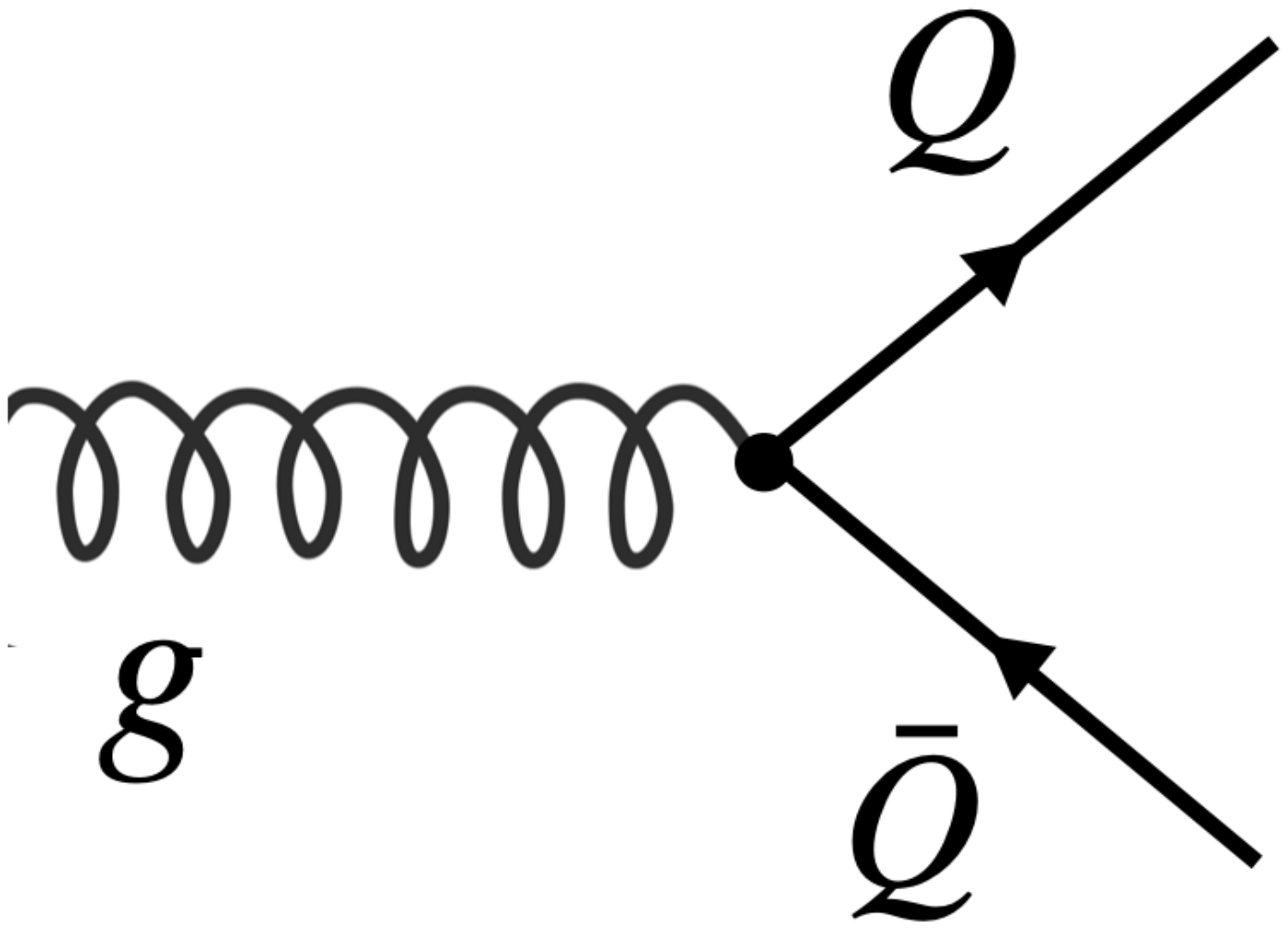} 
\caption{Elementary process of the gluon decay into quark anti-quark pair.}
\label{fig-1} 
\end{figure}

To verify the validity of this LLL approximation, we put our calculation a step forward, extending to the next Landau level. From the result, we could conclude that if the partons mainly distributed in the small x region and the heavy quark produced at low $p_T$ region with external magnetic field $\sim 10 GeV^2$ then the process could be well described at LLL. In the calculation, we found that the cross section of such an elementary process will encounter a divergence for $u-$ and $t-$ channel if we set the final quarks up to $n=1$ state. The inducement of this divergence is not from the threshold of the initial energy but the $on-shell$ of the internal quark, which means the Feynman diagram could be cut, and the sub-process is just the decay of the on shell gluon to quark anti-quark pair. In vacuum, the elementary process like Fig.~\ref{fig-1} is forbidden due to the energy-momentum conservation because the massless gluon has no rest mass.  However, with the background field, this process will no longer be forbidden any more because the strong magnetic field could compensate the energy. This process has been widely studied in the astro physics when they consider the QED processes on a magnetars. In heavy-ion collision, when considering the pre equilibrium stage with magnetic field, the gluon decay has barely contributed to the quark statistics.  In this proceeding paper, we will calculate and discuss this process under LLL.

The Feynman rules associated with quarks are controlled by Dirac equation
\begin{equation}
\label{Dirac}
\left[i\gamma^\mu \left(\partial_\mu+iqA_\mu\right)-m\right]\psi=0,
\end{equation}
where $m$ is the quark mass, $q$ the quark electric charge, and $A_\mu$ the electromagnetic potential. We consider an external magnetic field $B$ in the direction of $z$-axis and choose the Landau gauge with $A_0=0$ and ${\bm A}=Bx{\bm e}_y$. In this case, the momentum along the $x$-axis is not conserved. Taking into account the Landau energy levels for a fermion moving in an external magnetic field, the stationary solution of the Dirac spinor can be written as
\begin{equation}
\label{spinor}
\psi_{n,\sigma}^-(x,p) = e^{-ip\cdot x} u_{n,\sigma}({\bm x},p),\ \ \ \ \psi_{n,\sigma}^+(x,p) = e^{ip\cdot x} v_{n,\sigma}({\bm x},p)
\end{equation}
where the quantum number $n$ is Landau level, and the four-momentum is defined as $p_\mu = (\epsilon,0,p_y,p_z)$ with $p_y=aqB$ controlled by the center of gyration $a$. 

With the stationary solution of fermion field above, we could obtain the S-matrix of the decay process $g\to q\bar q$ as
\begin{equation}
\label{smatrix}
S_{fi} = -ig\int d^4x \bar\psi_-^{(\sigma_-)}(x)_{p_{z-},0,a_{-}}\gamma_\mu t^c A^{\mu}_c(x)\psi_+^{(\sigma_+)}(x)_{p_{z+},0,a_+}
\end{equation}
when we proceed our calculation under LLL, we could obtain 
\begin{eqnarray}
\label{smatrix1}
S_{fi}& =& t^c\frac{g}{2\sqrt{2\omega}L^{7/2}} e^{-\frac{i}{4}\omega \sin \theta [4 a_- \cos \phi + \lambda_B^2\omega \sin \theta(-i+\sin(2\phi))]}\nonumber\\
&\times& \epsilon_z \frac{p_{z-}(-p_{z-}+\omega\cos\theta)+(m+E_-)(m+E_+)}{\sqrt{m(m+E_-)}\sqrt{m(m+E_+)}}(2\pi)^3\delta_{fi}^3(E,p_y,p_z)
\end{eqnarray}
where $\omega \hat k = \omega(1,\sin\theta\cos\phi,\sin\theta\sin\phi,\cos\theta)$ is the momentum of the initial gluon and $\epsilon_z$ is the z-component of its polarization vector.  

The two independent polarization modes of gluon is $\mathcal O$ mode and $\mathcal E$ mode, with $\epsilon_{\mathcal O} = (-\cos\theta \cos\phi, -\cos\theta\sin\phi ,\sin \theta)$ and $\epsilon_{\mathcal E}=(\sin\phi, -\cos\phi,0)$. Since the S-matrix is proportional to $\epsilon_z$, only $\mathcal O$-mode remains. If here we set $\theta = \pi/2$, then $\epsilon_z = 1$ and $p_{z+} = -p_{z-},\ E_+=E_-=E$, we could obtain a simplified formulation
\begin{equation}
\label{smatrix2}
S_{fi} = t^c\frac{gm}{E\sqrt{2\omega}L^{7/2}} e^{-\frac{1}{4}\lambda_B^2\omega^2 + i \Phi} (2\pi)^3\delta_{fi}^3(E,p_y,p_z)
\end{equation}
where $\Phi$ is a phase term which will not contribute to the decay rate. And this S-matrix has the exact the same formula with ~\cite{Kostenko:2018cgv,Kostenko:2019was} except the color factor. 
The decay rate of the process $g\to Q\bar Q$($\theta = \pi/2$)
\begin{eqnarray}
\Gamma &=& g^2\sum_c\frac{1}{T}\int L\frac{dp_{z+}}{2\pi}\int L \frac{dp_{z-}}{2\pi}\int L\frac{da_{+}}{2\pi \lambda_B^2}\int L\frac{da_{-}}{2\pi\lambda_B^2}|S_{fi}|^2\nonumber\\
&=&g^2\frac{4q^2m^2}{12\pi E^2\omega \lambda_B^2}e^{-\lambda_B^2\omega^2/2}\int\frac{da_-}{L}\int dp_{z-}\delta(\omega-E_+-E_-)\nonumber\\
&=& g^2\frac{4qB m^2}{3\pi\omega^2\sqrt{\omega^2-4m^2}}e^{-\lambda_B^2\omega^2/2}
\end{eqnarray}

\begin{figure}[h]
\centering
\includegraphics[width=5.5cm]{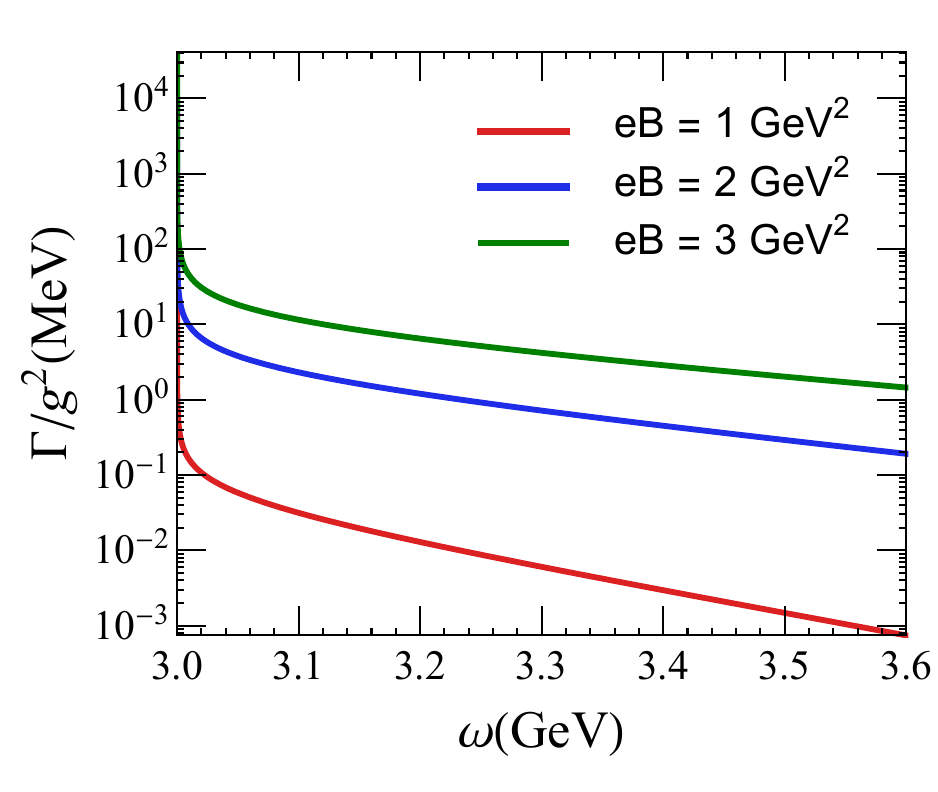} ~ \includegraphics[width=5.5cm]{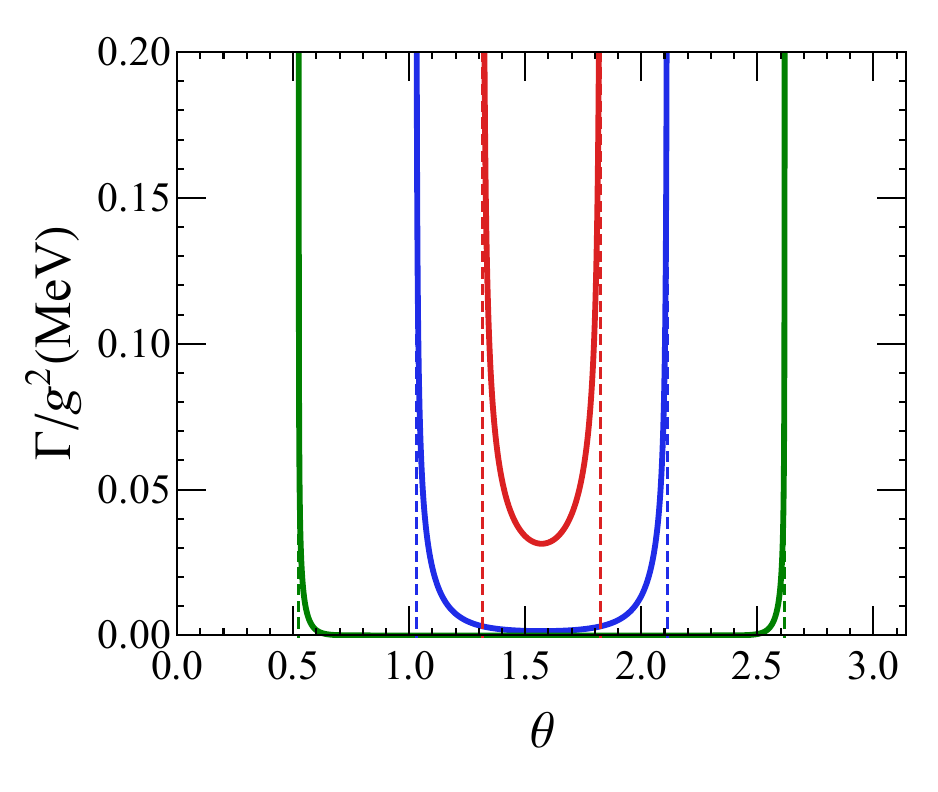} 
\caption{(Left)The decay rate of $g\to Q\bar Q$ with final quark mass $m=1.5$GeV and $\theta = \pi/2$ changes with initial gluon energy. (Right) The decay rate changes with polarization angle at $eB = 1~\rm GeV^2$ and different lines represent $\omega = 3.1~\rm GeV$(Red), $3.5~\rm GeV$(Blue) and $6.0~\rm GeV$(Green) respectively. Dashed lines are the thresholds of angle window determined by the energy conservation condition.}
\label{fig-2} 
\end{figure}

From the numerical result, we could see that the decay rate decreasing with incoming energy has the same property with the scattering cross section $gg\to Q\bar Q$, which has a threshold depending on the rest mass of final fermions and increasing with magnetic field. The difference appears when considering the polarization angle dependence. The energy conservation condition $\sqrt{p_{z+}^2+m^2}+\sqrt{p_{z-}^2+m^2}=\omega$ is satisfied only with polarization angle is around $\pi/2$ which means the gluon is moving perpendicular to the magnetic field especially when the gluon energy approaches threshold. And when the energy grows, the angle window becomes larger and along with the exponential decrease of the magnitude of decay width. On the contrary, when the incoming gluon is moving along the magnetic field, it is obvious to see that the energy conservation is violated and this process is forbidden. 

In summary, we derived the gluon decay rate under the external magnetic field in the strong field limit. This process is barely considered in the initial stage quark production since the magnetic field is believed to decay at a very fast speed when the system enters the QGP phase. From the result, we could conclude that the decay rate of gluon increases with the magnitude of magnetic field which means a stronger magnetic field can induce a faster decay of the gluon even if the gluon has no interaction with magnetic field. As our previous work mentioned, the LLL is more suitable for a lighter quark, and we believed that the light quark production will also be affected by this process.

{\bf Acknowledgement}: The work is supported by the NSFC grants No. 12075129, the Guangdong Major Project of Basic and Applied Basic Research No.2020B0301030008, the funding from the European Union’s Horizon 2020 research, and the innovation program under grant agreement No. 824093 (STRONG-2020).


\begin{thebibliography}{}
%
%



\bibitem{Kharzeev:2007jp}
D.~E. Kharzeev, L.~D. McLerran, and H.~J. Warringa, {\it {The Effects of
  topological charge change in heavy ion collisions: 'Event by event P and CP
  violation'}},  {\em Nucl. Phys. A} {\bf 803} (2008) 227--253,
  [{{\tt arXiv:0711.0950}}].



\bibitem{Voronyuk:2011jd}
V.~Voronyuk, V.~D. Toneev, W.~Cassing, E.~L. Bratkovskaya, V.~P. Konchakovski,
  and S.~A. Voloshin, {\it {(Electro-)Magnetic field evolution in relativistic
  heavy-ion collisions}},  {\em Phys. Rev. C} {\bf 83} (2011) 054911,
  [{{\tt arXiv:1103.4239}}].

\bibitem{Deng:2012pc}
W.-T. Deng and X.-G. Huang, {\it {Event-by-event generation of electromagnetic
  fields in heavy-ion collisions}},  {\em Phys. Rev. C} {\bf 85} (2012) 044907,
  [{{\tt arXiv:1201.5108}}].

\bibitem{Tuchin:2013ie}
K.~Tuchin, {\it {Particle production in strong electromagnetic fields in
  relativistic heavy-ion collisions}},  {\em Adv. High Energy Phys.} {\bf 2013}
  (2013) 490495, [{{\tt
  arXiv:1301.0099}}].






\bibitem{ALICE:2015mzu}
{\bf ALICE} Collaboration, J.~Adam et~al., {\it {Measurement of an excess in
  the yield of $J/\psi$ at very low $p_{\rm T}$ in Pb-Pb collisions at
  $\sqrt{s_{\rm NN}}$ = 2.76 TeV}},  {\em Phys. Rev. Lett.} {\bf 116} (2016)
  222301, [{{\tt arXiv:1509.08802}}].




\bibitem{Das:2016cwd}
S.~K. Das, S.~Plumari, S.~Chatterjee, J.~Alam, F.~Scardina, and V.~Greco, {\it
  {Directed Flow of Charm Quarks as a Witness of the Initial Strong Magnetic
  Field in Ultra-Relativistic Heavy Ion Collisions}},  {\em Phys. Lett. B} {\bf
  768} (2017) 260--264, [{{\tt
  arXiv:1608.02231}}].





\bibitem{Guo:2019joy}
Y.~Guo, S.~Shi, S.~Feng, and J.~Liao, {\it {Magnetic Field Induced Polarization
  Difference between Hyperons and Anti-hyperons}},  {\em Phys. Lett. B} {\bf
  798} (2019) 134929, [{{\tt
  arXiv:1905.12613}}].

\bibitem{Chen:2024lmp}
S.~Chen, J.~Zhao and P.~Zhuang,
``Heavy flavor production under a strong magnetic field,''
[arXiv:2401.17559 [hep-ph]].


\bibitem{Wang:2021oqq}
Z.~Wang, J.~Zhao, C.~Greiner, Z.~Xu, and P.~Zhuang, {\it {Incomplete
  electromagnetic response of hot QCD matter}},  {\em Phys. Rev. C} {\bf 105}
  (2022) L041901, [{{\tt
  arXiv:2110.14302}}].

\bibitem{Yan:2021zjc}
L.~Yan and X.-G. Huang, {\it {Dynamical evolution of magnetic field in the
  pre-equilibrium quark-gluon plasma}},
 {{\tt arXiv:2104.00831}}.

\bibitem{Chen:2021nxs}
Y.~Chen, X.-L. Sheng, and G.-L. Ma, {\it {Electromagnetic fields from the
  extended Kharzeev-McLerran-Warringa model in relativistic heavy-ion
  collisions}},  {\em Nucl. Phys. A} {\bf 1011} (2021) 122199,  [{{\tt arXiv:2101.09845}}].
  
  \bibitem{Kostenko:2018cgv}
A.~Kostenko and C.~Thompson, {\it {QED Phenomena in an Ultrastrong Magnetic
  Field. I. Electron\textendash{}Photon Scattering, Pair Creation, and
  Annihilation}},  {\em Astrophys. J.} {\bf 869} (2018) 44,
  [{{\tt arXiv:1904.03324}}].

\bibitem{Kostenko:2019was}
A.~Kostenko and C.~Thompson, {\it {QED Phenomena in an Ultrastrong Magnetic
  Field. II. Electron-Positron Scattering, $e^\pm$-Ion Scattering, and
  Relativistic Bremsstrahlung}},  {\em Astrophys. J.} {\bf 875} (2019) 23,
  [{{\tt arXiv:1904.03325}}].





\end{thebibliography}
%
%

\end{document}